\documentclass[pre,twocolumn,showpacs]{revtex4-1}
\usepackage{amsmath,amssymb,latexsym}
\usepackage{rotating}
\usepackage{color}
\usepackage[utf8]{inputenc}
\usepackage{graphicx} 
\graphicspath{{./figures}}
\usepackage{epstopdf}
\usepackage{xcolor}
\usepackage{ulem}

\usepackage[colorlinks = true,
            linkcolor = blue,
            urlcolor  = blue,
            citecolor = blue,
            anchorcolor = blue]{hyperref}

\begin{document}

\title{The finite-$T$ Lorentz number and the thermal conductivity.\\
Aluminum and carbon conductivities
from ambient to  millions of degrees Kelvin}

\normalsize

\author
{
 M.W.C. Dharma-wardana}
\email[Email address:\ ]{chandre.dharma-wardana@nrc-cnrc.gc.ca}
\affiliation{
National Research Council of Canada, Ottawa, Canada, K1A 0R6,
\& Universit\'{e} de Montreal, Montreal, Canada, H3C 3J7
}

\date{\today}
\begin{abstract}
Theoretical prediction of the thermal conductivity $\kappa$ of metal-like
 electron-ion systems would be greatly simplified if
a convenient generalization of the Lorentz number $L_N$ for arbitrary
temperatures ($T$) and densities were available. Such calculations are needed
in astrophysics, high-energy-density physics, semiconductor physics
 as well as in materials science. We present a finite-$T$ form of $L_N(T)$, 
expressed in terms of elementary Fermi integrals. It is a universal function
of $t=T/E_F$, where $E_F$ is the Fermi energy of the electrons. A convenient
 four-parameter fit to $L_N(t)$ for $t=0-\infty$ further simplifies
 the applications.
 The effect of electron-electron interactions is also briefly discussed.
Calculations for $L_N(t)$ and thermal conductivities $\kappa$ for Al and C are
presented at several compressions and into the million-Kelvin range.
Experimental isobaric conductivities for Al just above the meting point,
and isochoric conductivities for Al and C from
available density-functional theory
 simulations and average-atom calculations are used as comparisons.  
\end{abstract}
\maketitle

{\it Introduction-} Modern studies in materials science, be it for hydrogen
 storage or astrophysical 
and warm dense matter (WDM) applications, seek to predict the relevant target physical
properties prior to costly and time-consuming experimentation. In many cases of WDM, and
most astrophysical systems, experimentation is usually impossible except for narrow ranges
of the phase diagram. These systems straddle intermediate regimes of  matter 
extending from cold solids to hot plasmas, at various densities and states of ionization
\cite{Nature24,Stanek24,DW-yul22,GaffneyHDP18, Grabowski20, Hungary16, cdw-cpp15, NgIJQC12}.
The same issues arise in high-energy-density  applications where energy is deposited
in femto-second  time-scales producing multi-temperature 
quasi-equilibria~\cite{Neder22,KandTrickey2020,NgIJQC12,CDW-cm-91,ELR98} 
where temperature-relaxation rates are of interest.
The electrons and holes in semi-conductor nanostructures also behave like WDM
systems due to the small effective masses of the charge carriers~\cite{lfc1-dw19}.

While crystalline solids are defined by a few atoms in a unit cell,
 systems above the melting point pose challenges for traditional $N$-atom
 density-functional theory (DFT) and  molecular-dynamics (MD) based simulations where $N$
 must be made as large as possible. 
The ions are usually modeled
 as classical particles,  where as the electrons are treated quantum
 mechanically \cite{stanton2018multiscale}. If finite temperatures are to be treated,
 then the number of basis functions needed for the quantum calculation
becomes prohibitively high, when
 the temperature $T$ 
(in energy units) exceeds the Fermi energy $E_F$. 
This $N$-atom DFT-MD approach will be referred to as the quantum molecular dynamics (QMD) method. Its current
 implementations do not usually include finite-$T$ electron exchange and correlation
 (XC) effects although finite-$T$ functionals are
 available~\cite{PDWXC84,PDWXC,KSDT14,GDS17,Dornheim18,Karasiev19}. In QMD,  $N$-ions
 are explicitly simulated while the electron-electron interaction is reduced
 to a one-electron problem.

 Instead of QMD we use the neutral-pseudo-atom model (NPA),
  a rigorous DFT approach~\cite{DWP82,DW-yul22} that eliminates explicit
 simulation of  $N$-ions  by invoking an ion-ion XC functional (in addition to the
e-e XC functional), to incorporate multi-ion effects.
In the NPA, the free-electron density $n_f(r)$ and the mean free-electron density $\bar{n}$
are available from the Kohn-Sham calculation for a single nucleus immersed in the
 fluid defined by its self-consistently determined one-body  densities
  $n(r)$ and $\rho(r)$ for electrons and ions respectively.  The density increment
 $\Delta n_f(k)=n_f(k)-\bar{n}$ serves to define a weak pseudopotential $U_{ei}(k)$ 
that may be used in evaluating the ion-ion $S(k)$,
and in transport calculations (static fields).
The pseudopotential subsumes the internal structure of the ion and describes
 a rigid scatterer. The T-matrix constructed from the phase-shifts
 of the NPA calculation can also be used in the Ziman-Evens formula and can be sensitive
to the internal structure of the ion. 
  
Calculations of {\it static} transport quantities in QMD are complex, indirect and  mired in
many assumptions. In QMD, Kohn-Sham electronic states and energies are
 generated for a succession of {\it fixed} configurations of $N$-ions used in the
 simulation, thus neglecting ion-dynamical effects and coupled-mode effects that are
 important near low-frequencies $\omega\to 0$. Typically $64\le N \le 512$. The dynamic
 conductivity of the fluid, $\sigma(\omega)$, is evaluated as a thermal average
 of the crystalline $\sigma_c(\omega)$ values of `all'  periodic solids constructed from
 the $N$-ions in the QMD simulation. The Kubo-Greenwood (KG)
 formula~\cite{Dejarlais02,Recou05} is used for $\sigma(\omega)$. The local-field
corrections arising from the crystalline band structure~\cite{Adler62} are ignored, and  
the calculated KG-$\sigma(\omega)$, lacking in ion dynamics diverges as
$\omega\to 0$. Hence at this stage a Drude model or other
 model~\cite{Johnson09, KuchievJohnson08} that assumes a rigid internal structure
 for the ions, and a scattering potential
that remains unchanged over the long-time scales
of the $\omega\to 0$ collisional limit are assumed to extrapolate $\sigma(\omega)$ to $\sigma(0)$.
 The evaluation of $\kappa$ proceeds via the kinetic
 coefficients of semiclassical transport theory~\cite{Recou05,AshMer76}. 

 The Lorentz number $L_N$ is traditionally evaluated from independently determined
$\sigma$ and $\kappa$. The temperature $T_k$  is usually expressed in
 Kelvin. 
\begin{equation}
L_N(T_k)=\frac{\kappa}{\sigma(0)T_k} =\frac{\kappa}{\sigma T}\left[\frac{k_B}{e}\right]^2.
\end{equation}
We use Hartree atomic units, with $\hbar=m_e=|e|=1$, and $T$ such that the
Boltzmann constant $k_B=1$. The electron Wigner-Seitz radius $r_s$ is given by
 $r_s=(3/4\pi\bar{n})^{1/3}$. The number of free electrons per ion, viz.,
 $\bar{Z}$ is $\bar{n}/\bar{\rho}$, where $\bar{\rho}$ is the mean ion density. 

The Lorentz number of ideal electrons  scattering
from a system of heavy ions (Lorentz gas) at $T=0$, and for $T\to\infty$ (classical limit)
 is well known. Here we derive a result  for arbitrary degeneracies and densities
 in terms of elementary Fermi integrals. 
Then $L_N(T,r_s)$ is found to be a universal function of $t=T/E_F$ only, 
where $E_F$ is the Fermi energy. Hence, given an estimate of $\sigma$, 
the thermal conductivity $\kappa$ is immediately known via $L_N(t)$. At this
stage electron-electron interactions and ion-embedding effects are neglected.

Electron-electron interactions are included in traditional transport theories by
additional corrections via the Boltzmann equation~\cite{SpHarm53}, via perturbation
 corrections to kinetic equations and distribution functions
 etc~\cite{Shtermin2006,Reinholz2015,Desjarlais17}.
 However, if phase-shifted plane waves are  eigenstates of electrons, no
contributions from e-e interactions  arise. Constructing higher-order theories
to include e-e interactions  within conserving approximations is very difficult,
 especially at finite-$T$~\cite{RichAsh94}. DFT solves the problem 
of e-e interactions by mapping interacting electrons to non-interacting Kohn-Sham electrons
experiencing just a one-body XC-potential.  Hence no further e-e corrections are
 needed if NPA generated  $U_{ei}(k), S(k)$, T-matrices etc., are 
used in transport calculations (see SM~\cite{SM}) as they already include
 finite-$T$ XC effects.
   
Explicit isochoric calculations for liquid Al ($l$-Al) at $\bar{\rho}$=2.7 g/cm$^3$,
 2.35 g/cm$^3$ are presented, while isobaric calculations are presented for the
 densities studied experimentally by Leitner et al~\cite{Leitner17}. Calculations
 of the Lorentz number, electrical and
thermal conductivities for $l$-C at $\bar{\rho}$=10 g/cm$^3$ and at the diamond-like
density of 3.6 g/cm$^3$ (see SM) are also presented and compared with any
 available published  data. 
\begin{figure}[t]
\includegraphics[width=0.96\columnwidth]{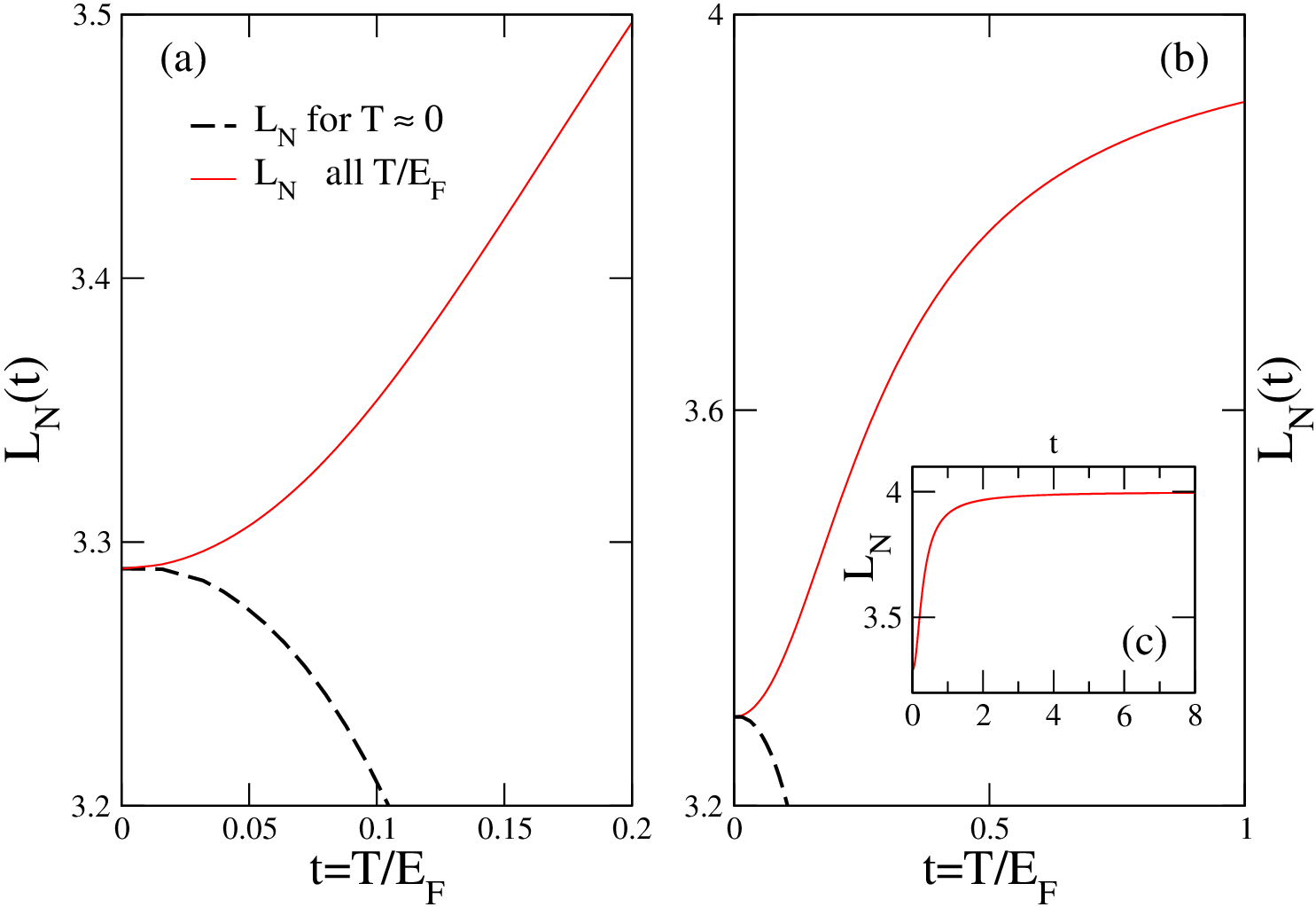}
\caption{\label{lorfit.fig}(Color online) (a) The universal Lorentz number $L_N(t)$ 
Eq.~\ref{lorT.eqn}, fitted to Eq.~\ref{Lorfit.eqn} is displayed by the
continuous red line for $t>0.2$.
 The $T\sim 0$ approximation, Eq.~\ref{kappaT0.eqn} is
shown as a dashed black line. (b) The same as the left panel, now for $t\le 1$.
 The inset (c) shows $L_N(t)$ from the quantum limit $\pi^2/3$
to the classical (Landau-Spitzer) limit of 4.}  
\end{figure}

{\it The Lorentz number and the conductivities-} In the semi-classical theory of transport,
the kinetic coefficients $\mathcal{L}^{(\alpha)}$ are used to express
the conductivities. Their detailed definitions and reduction to simple forms within
a Lorentz-plasma model are given in SM~\cite{SM}. Then we have
\begin{eqnarray}
\sigma &=&\mathcal{L}(0),\;
\kappa =\left[\mathcal{L}(2)-(\mathcal{L}^{(1)})^2/\mathcal{L}^0\right]/T\\
L_N &=& \kappa/(T\sigma).
\end{eqnarray}
In SM we develop the reduced form
\begin{equation}
\mathcal{L}^{(\alpha)}=\langle\frac{1}{3}(\epsilon-\mu)^{\alpha} 
 \tau_{ie}(\epsilon)(2\epsilon)^{3/2}\rangle.
\end{equation}
The mean value $\langle\cdots\rangle$ is defined by:
\begin{eqnarray}
\langle g(\epsilon)\rangle &=&-\int_0^\infty d\epsilon\frac{f(\epsilon)}{d\epsilon} g(\epsilon)\\
f(\epsilon)&=&1/\left[1+\exp(\epsilon-\mu)/T\right].
\end{eqnarray}
The relaxation time $\tau_{ei}$ can be evaluated via the NPA using the static ion-ion
 structure-factor $S(k)$ and the cross section $\Sigma(\vec{k},\vec{q})$. The latter
 is for the scattering of an electron of
 momentum $\vec{k}$ with a momentum transfer $\vec{q}$ ~\cite{Thermophys99}. In elastic 
collisions (with the internal structure of the ion held invariant), a weak pseudopotential
can be constructed to describe the scattering~\cite{DW-yul22}. 
The energy dependent $\tau_{ie}(\epsilon)$ can be written as
\begin{equation}
\tau_{ie}(\epsilon)=C_{\tau}\epsilon^{3/2}/C_{lg}.
\end{equation}
Here the numerical coefficient $C_\tau$ and the Coulomb logarithm $C_{lg}$ can be
matched, as desired, to a sophisticated calculation of $\tau_{ie}$ or to
 a simple Landau-Spitzer form. 
Then it is shown in SM ~\cite{SM} that the $L_N(t)$ is given by:
\begin{eqnarray}
\label{lorT.eqn}
L_{\rm N}(t)&=&\frac{5I_4(\eta)}{3I_2(\eta)}-\left[\frac{4I_3(\eta)}{3I_2(\eta)}\right]^2,\; t> 0 \\
            &=&\pi^2/3, \;  t\sim  0 \\
        I_n(\eta) &=& \int \frac{dx x^n}{1+\exp(x-\eta)},\; \eta=\mu/T.
\end{eqnarray}
These results can be fitted to the formula
\begin{equation}
\label{Lorfit.eqn}
L_{\rm N}(t)=\frac{a_0+a_1t+a_2t^2+a_3t^3+4t^4}{1+b_2t^2+t^4}.
\end{equation}
The fit coefficients  $a_0,a_1,a_2,a_3$ are $\pi^2/3$, -0.0666174, 43.7111, 0.00994066
respectively, while $b_2=11.010134$. This result does not include effects of ion-embedding
in the electron fluid, or XC-effects. For high-density electron systems like aluminum
 or carbon at the densities studied here, ion-embedding effects mainly result in a
rigid lowering of the continuum. When electron XC-effects are included, $a_0$ is
modified from 3.29 to 3.328 near $T=0$ for Al at 2.7 g/cm$^3$, and
to 3.194 for carbon at 10 g/cm$^3$, but these are countered by other contributions. These
many-body effects will be considered in a separate study.

The $t\simeq 0$ limiting form is obtained from
the following.
\begin{equation}
\label{kappaT0.eqn}
\kappa =(1/3)C_e<V^2>\tau_{ie}, \;\;  t\sim 0
\end{equation}
Here $C_e$ is the electron specific heat per particle, while  $<V^2>$
is an electron mean square-velocity evaluated within the
scattering region $ E_F\pm T$ in $k$-space enclosing  the
Fermi energy. We compare this limiting form for $L_N(t\sim 0)$
 with the fit form, Eq.~\ref{Lorfit.eqn}, numerical equivalent to
 the analytic form, Eq.~\ref{lorT.eqn}
in Fig.~\ref{lorfit.fig}. The same curve $L_N(t)$ is obtained for all
$r_s,T$ given as a universal function of $T=T/E_F$. This universality is lost
when XC-effects, embedding effects~\cite{eos95}
 are included. For instance, the modification of
 the specific heats can be calculated from the second derivative of the
total electron free energy that includes finite-$T$ XC-effects~\cite{Karasiev19}.

{\it Thermopower and the Seebeck coefficient-} The thermoelectric coefficient (Seebeck 
coefficient) $\theta(t)$ can be expressed as:
\begin{equation}
\label{theta.eqn}
\theta(t)=-\frac{1}{eT}\frac{\mathcal{L}^{(1)}}{\mathcal{L}^{(0)}}
        =\left(\frac{4I_3}{3I_2}-\eta\right),\; t>0.
\end{equation}
The above equation is not applicable at $T\simeq 0$. In the classical limit
 Eq.~\ref{theta.eqn} takes the value $(4-\eta)$.

\begin{figure}[t]
\includegraphics[width=0.96\columnwidth]{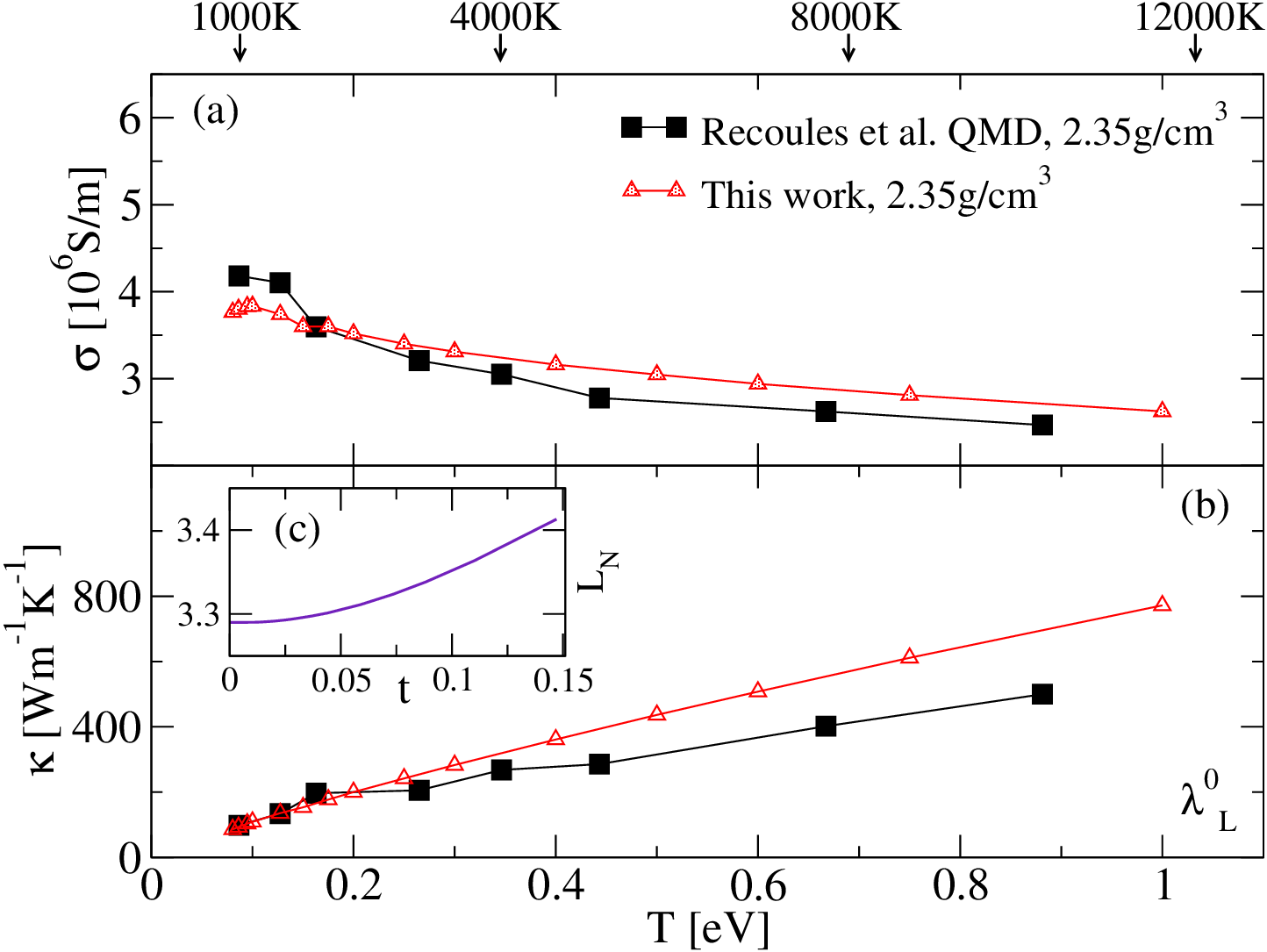}
\caption{\label{Recoul.fig}(Color online) (a) The electrical conductivity $\sigma_z$ for $l$-Al
at 2.35 g/cm$^3$ from the NPA-pseudopotentials and the Ziman formula. The QMD-KG results of Recoules
et al~\cite{Recou05} are at 2.35 g/cm$^3$. Panel (b) shows the corresponding $\kappa$
The inset (c) shows $L_N(t), t=T/E_F$ Eq.~\ref{Lorfit.eqn} used in calculating our $\kappa$.}
\end{figure}

{\it Thermal conductivities of Al and carbon-} Recoules et al~\cite{Recou05} have calculated
the electrical conductivity $\sigma$ using QMD-KG and the corresponding $\kappa$ via the 
kinetic coefficients for isochoric $l$-Al at the density 2.35 g/cm$^3$ from 1000K (0.0862 eV) to
 10,000K (0.862 eV). Their results, as well as our results for $\sigma$ evaluated from the
pseudopotential, and the corresponding $\kappa$
 are shown in Fig.~\ref{Recoul.fig}.

\begin{figure}[t]
\includegraphics[width=0.96\columnwidth]{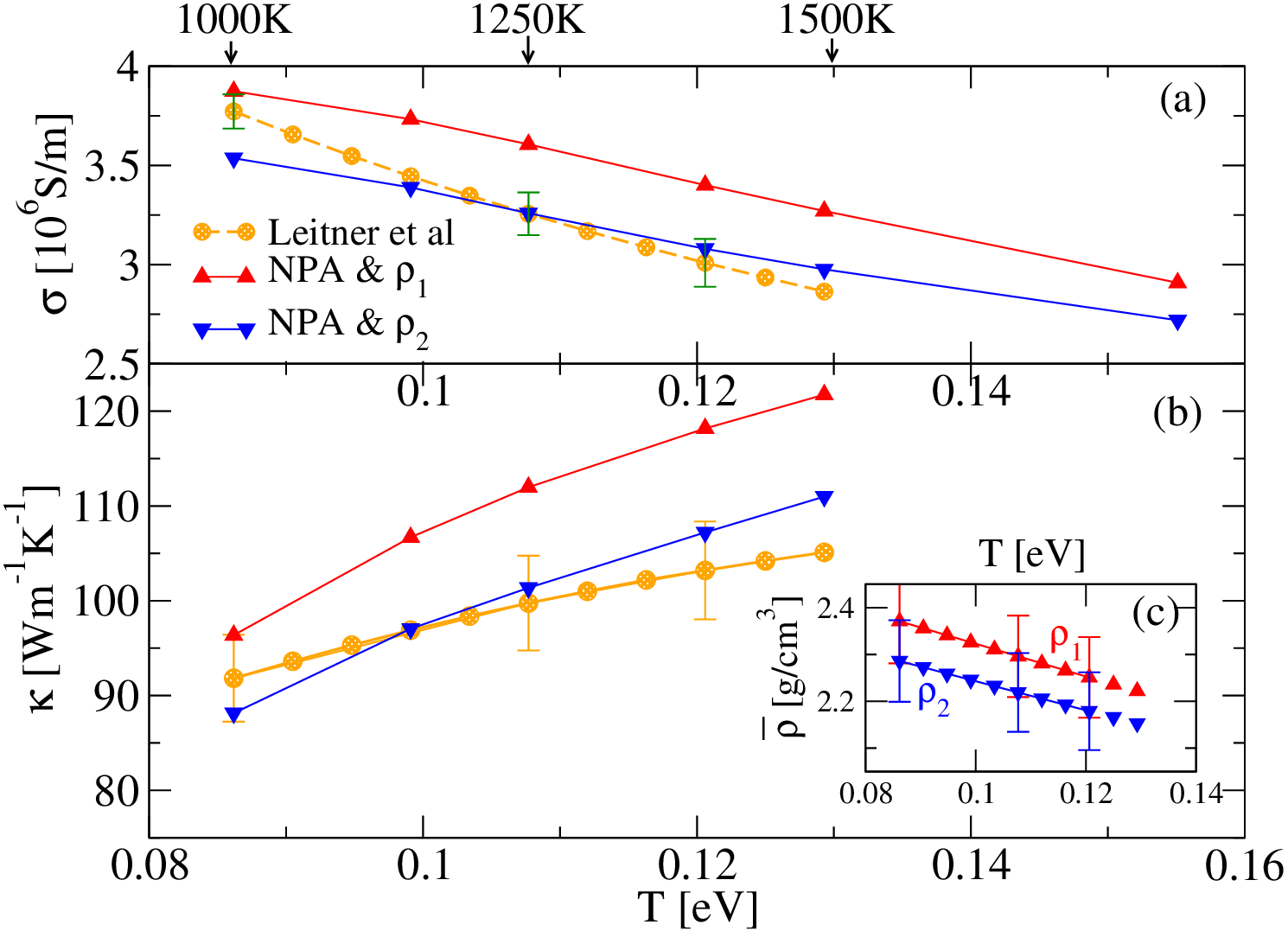}
\caption{\label{Leitner.fig}(Color online) (a) The electrical conductivity $\sigma_z$ for $l$-Al
at isobaric densities $\rho_1,\rho_2$  from the NPA-pseudopotentials and the Ziman formula, compared
 with the experimental results of Leitner et al~\cite{Leitner17}. Panel (b) shows the
 corresponding $\kappa$ calculated using the  $L_N$ of our theory, as well as the estimates in 
Ref.~\cite{Leitner17}. The inset (c) shows the two sets of densities $\rho_1,\rho_2$ that
are upper and lower bounds to the actual isobaric density $\rho(T)$. The better estimate of $\rho(T)$ is 
possibly $\rho_{2}$.}
\end{figure}

Direct experimental results for $\kappa$ for liquid metals do not seem to be available. Leitner
 et al.~\cite{Leitner17} have reported accurate experimental isobaric electrical conductivities
for $l$-Al  near the melting point and {\it evaluated} $\kappa$ using the ideal Lorentz  number
 which is practically identical to the estimate from our Eq.~\ref{Lorfit.eqn} as well. 
Their isobaric density measurement is constrained to  upper and  lower estimates $\rho_1(T)$ and
$\rho_2(T)$. We calculate $\sigma,\kappa$ for the upper and lower sets of densities and compare
them (see Fig.~\ref{Leitner.fig})  with the results of Leitner et al~\cite{Leitner17}. Our theoretical
analysis supports their lower density estimate $\rho_2$.

{\it High-temperature results-} While experimental $\sigma$ data are available near
 the melting point of common metals, data at higher-$T$ data are unavailable. Hence we compare our results
 with other theoretical results. While the $\sigma_z$ calculated from a pseudopotential is for
 robust ions of charge $\bar{Z}$ where the scattering electron does not interact with the weakly-bound
electronic structure of the ion, the T-matrix approach~\cite{Evens73} includes such interactions,
 as discussed further  in SM~\cite{SM}. Different approaches differ in the manner they deal with
 such `hopping electrons' associated with clusters of ions~\cite{hop1992} or an individual average
 ion, and in how they handle strong-scattering effects.
 Hence the $\sigma$ calculated  by the Ziman formula  using a pseudopotential based on $\bar{Z}$,
 denoted here by $\sigma_z$, from a T-matrix, viz $\sigma_{tmx}$
 or from the K-G approach may differ for ions with loose internal structure and in the treatment of
 strong-scattering.

\begin{figure}[t]
\includegraphics[width=0.96\columnwidth]{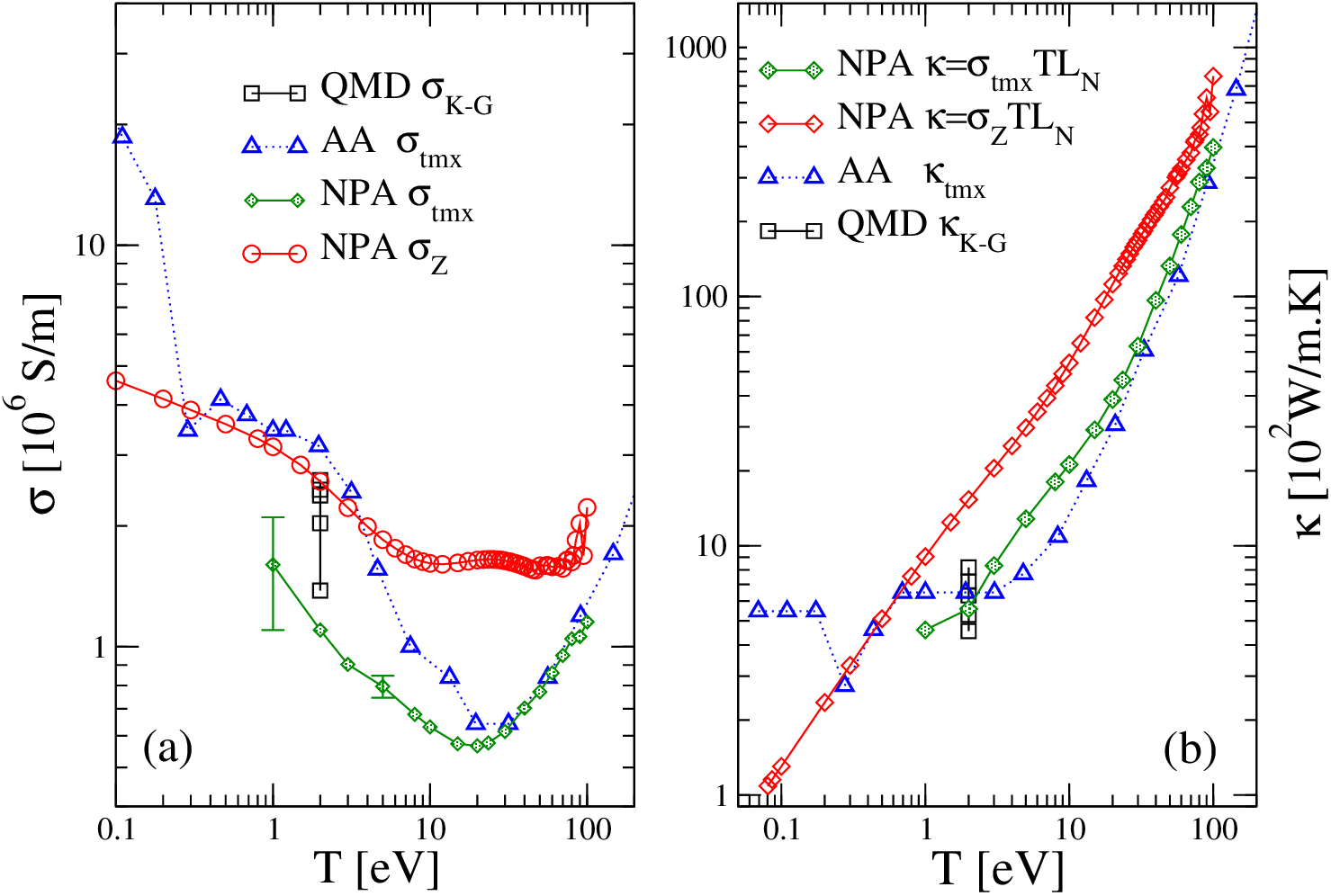}
\caption{\label{sig2p7Comp.fig}(Color online) (a) The isochoric conductivities 
$\sigma_z,\sigma_{tmx}$  calculated via the Ziman formula with NPA inputs, for
 $l$-aluminum, compared with selected  AA and all QMD results as reported
 in Ref.~\cite{Stanek24}. (b) The NPA-$\kappa$ obtained via
 Eq.\ref{Lorfit.eqn}, as well as selected thermal conductivities $\kappa$ reported
 in Ref.~\cite{Stanek24}.}
\end{figure}

\begin{figure}[b]
\includegraphics[width=0.96\columnwidth]{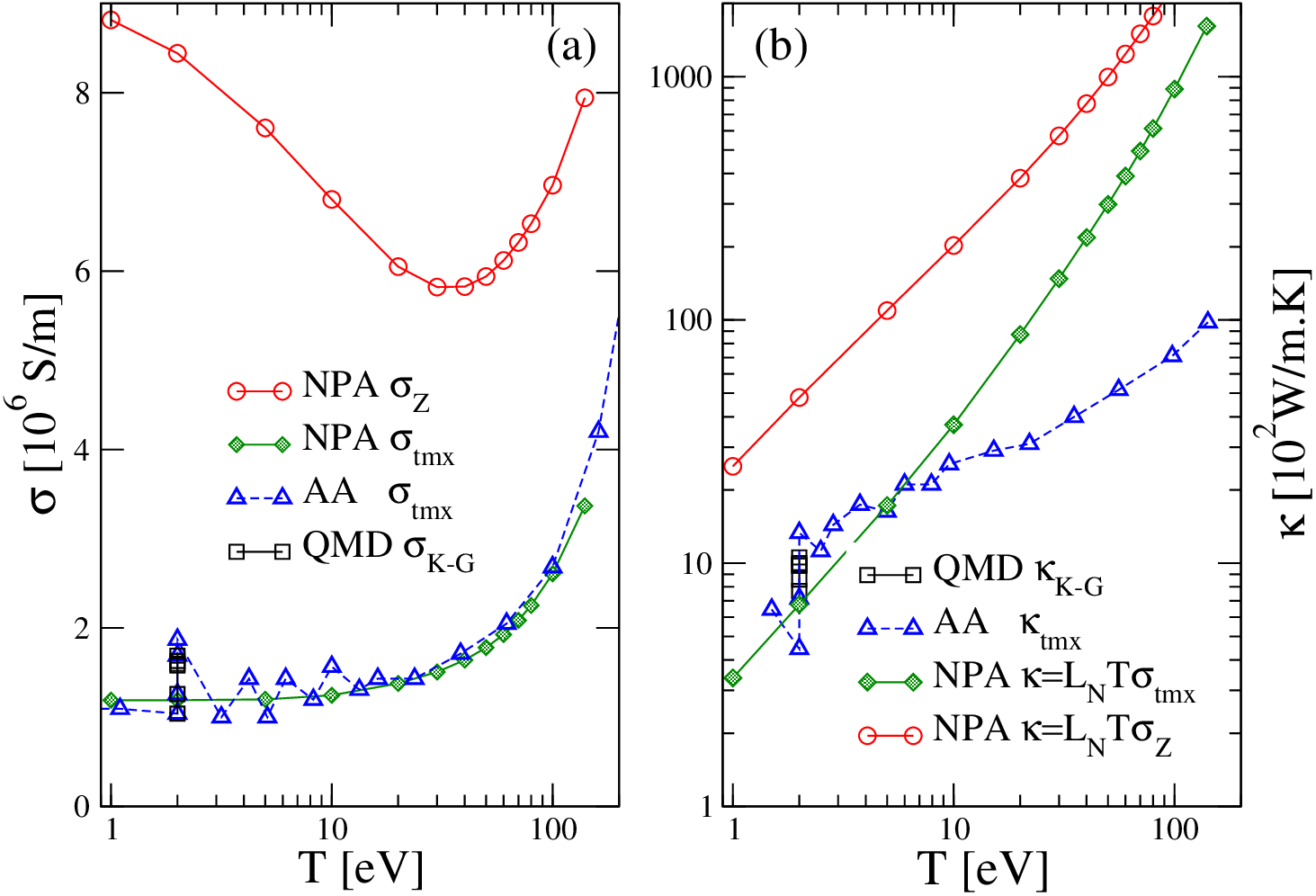}
\caption{\label{sig10Comp.fig}(Color online) Results for $l$-carbon as
 in Fig.~\ref{sig2p7Comp.fig}.
}
\end{figure}

Given a $\sigma$, it appears  that a good approximation to the corresponding $\kappa$
 can be obtained via the finite-$T$ Lorentz number, viz., Eqs.~\ref{lorT.eqn},\ref{Lorfit.eqn}.
 In the following and (in SM, for carbon at 3.6 g/cm$^3$), we present results for
 isochoric $\sigma_z,\kappa$ for
$l$- aluminum at 2.7 g/cm$^3$ and for $l$-carbon at 10 g/cm$3$, in Figs.~\ref{sig2p7Comp.fig},
~\ref{sig10Comp.fig} respectively.

{\it Discussion-} The Lorentz number and the Hall coefficient are two quantities that
 are relatively independent of the details of the material and
 the microscopic interactions included in $\sigma$ and $\kappa$. Thus,
even a minimal model of the interactions may be expected to provide a good evaluation of
  $L_N$. This is already seen in the quantum and classical limits of $L_N$ which
 are $\pi^2/3$ and 4 respectively. We show in SM that even at intermediate degeneracies,
 most of the complexities in $\mathcal{L}^{(\alpha)}$ factors cancel out in evaluating
  $L_N$. Hence our simplified evaluation of $L_N$ is likely to have a larger regime of
 validity than expected.

The good agreement of the Ziman conductivity $\sigma_z$ based on a pseudopotential,
 in the low-$T$ regime as seen in Figs.~\ref{Recoul.fig}, \ref{Leitner.fig} shows
 that the assumption of weak scattering from an Al$^{3+}$ ion with a rigid core is
 valid  up to $T \sim 1$ eV for Al. As $T$ increases, $\bar{Z}$ begins to exceed 3 and acquires
 fractional values, e.g., 3.25 etc. Then hopping-electron distributions emerge and the
 calculated $\sigma$ depends on how partially localized charges are dealt with in
 different conductivity  models~\cite{LM84,PDW-Res87,eos95,Rosznyai08,Johnson09,Filinov09,
 SternZbar07,Murillo13,StaSauDaligHam14,Pain2023}.  However, the $\sigma_z$ model provides a
 rigorous estimate of the {\it elastic scattering} contribution
 from ions whose internal structures remain rigid. If the phase shifts from the NPA
 are used in  a T-matrix calculation, denoted here as
 $\sigma_{tmx}$~\cite{SM,Thermophys99},
 the corresponding lower conductivity  of Al approaches that of other T-matrix models,
 both for $\sigma$ and $\kappa$,  with the latter evaluated here using our finite-$T$
 Lorentz number. For $T>E_F$, the NPA becomes increasingly
like an AA model in that the free-electron density increment $\Delta n_f(r)$
 is contained  within the
Wigner-Seitz  ion sphere. Hence the good agreement in $\sigma_{tmx}$ between the
NPA and AA results for $T>E_F$ found for $l$-Al is not surprising. The agreement between
$\kappa$ values shows that the $L_N$ used here for evaluating the NPA-$\kappa$ is
 trustworthy. For $l$-carbon at 10 g/cm$^3$, $E_F\simeq 60$ eV. Hence the
 agreement in $\sigma_{tmx}$ between NPA and AA even in the $t<1$ range
is interesting, but the AA-estimates of $\kappa$ for $T>10$ eV and from our calculations
disagree, unexpectedly. Further result using the  NPA and $L_N(t)$ equation are given
 in the supplemental material~\cite{SM}.   
%

\end{document}


\title{Supplemental Material for:\\
The finite-$T$ Lorentz number and the thermal conductivity.\\
Aluminum and carbon conductivities
from ambient to  millions of degrees Kelvin}

\normalsize

\author
{
 M.W.C. Dharma-wardana}
\email[Email address:\ ]{chandre.dharma-wardana@nrc-cnrc.gc.ca}
\affiliation{
National Research Council of Canada, Ottawa, Canada, K1A 0R6,
\& Universite de Montreal, Montreal, Canada, H3C 3J7
}

\date{\today}
\begin{abstract}
This supplemental material (SM) covers the following topics:\\
(i) Derivation of the finite-$T$ form of the Lorentz number $L_n(t)$\\
(ii) Electron-electron interactions in DFT and in conventional approaches.\\
(iii) The electrical conductivity from Pseudopotentials and from
the T-matrix form of the scattering cross section.\\ 
(iv) Results for $l$-Al at 2.7 g/cm$^3$, $l$-carbon at 10 g/cm$^3$, and
at the ``diamond-like'' density of 3.6 g/cm$^3$.
\end{abstract}
\maketitle

\section{Derivation of the finite-$T$ form of the Lorentz number} 
We use Hartree atomic units, with $\hbar=m_e=|e|=1$, and $T$ such that the
Boltzmann constant $k_B=1$.  The symbols defined in the main text are
also used in this supplemental material without further definition unless
additional clarification is needed.
 
In the following we use the following normalization of the Fermi function:
\begin{eqnarray}
\bar{n} &=&\int \frac{d\vec{k}}{4\pi^3}f(k)\\
  f(k)&=&f(\epsilon)=1/\left[1+\exp(\epsilon-\mu)/T\right]\\
  \epsilon&=&k^2/2.  
\end{eqnarray}
Then the classical limit is given by
\begin{eqnarray}
f(e)&=&\exp(\mu/T)\exp(-\epsilon/T)\\
e^{(\mu/T)}&=&\frac{\bar{n}}{2}\left[\frac{2\pi}{T}\right]^{3/2}.
\end{eqnarray}
The above normalization is commonly used~\cite{AshMer76}, but differs from the usage in
 some standard texts~\cite{LLvol10}. Furthermore, in the NPA, the volume occupied by the
 free electrons is {\it not} the ionic Wigner-Seitz sphere of radius 
$r_{ws}=\{3/4\pi\bar{\rho}\}^{1/3}$, but
an infinitely large volume, approximated by a volume of radius $R_c$ of
 the ``correlation sphere'' of the fluid. It is such that, given a nucleus placed
 at the origin of the correlation sphere,
all pair-distribution functions $g_{ss'}(r)$, where the species $s$ or $s'$ may be
 electrons or ions, have decayed to unity when $r \to R_c$. For $T<E_F$, usually $R_c\sim
10r_{ws}$, while for higher $T$ we have used $R_c\sim 5r_{ws}$.   

The non-interacting electrons (i.e., Kohn-Sham electrons) populating the correlation sphere
 take the noninteracting value $\mu^0$ for its chemical potential,
as required by DFT. This model is discussed in more detail in Refs.~\cite{DWP82,eos95}.
Average-atom models that confine the free electrons to within the ionic
Wigner-Seitz sphere become
similar to our NPA model for $T$ sufficiently large, such that $g_{ie}(r)=n(r)/\bar{n}$
 has already
 decayed to unity as $r\to r_{ws}$.

We define the kinetic coefficients $\mathcal{L}^{\alpha}$ as in Ashcroft and Mermin,
 Chapter 13~\cite{AshMer76}. We use the notation $\langle\cdots\rangle$ used in
 the main text to
indicate averaging over $-df(\epsilon)/d\epsilon$. Then, 
\begin{eqnarray}
\mathcal{L}^{\alpha}&=&\langle (\epsilon-\mu)^{\alpha}\tau(\epsilon)I(\epsilon)\rangle\\
I(\epsilon)&=&\int \frac{d\vec{k}}{4\pi^3}\delta(\epsilon-\epsilon(\vec{k}))
\vec{V}_{\vec{k}}\vec{V}_{\vec{k}}. 
\end{eqnarray}
As we are considering a uniform fluid with $\epsilon(\vec{k})=k^2/2$,
 $\vec{V}\vec{V}=(1/3)V^2$, the above equation can be written as:
\begin{eqnarray}
\mathcal{L}^{\alpha}&=&C_0
\langle (\epsilon-\mu)^{\alpha}\epsilon^{3/2}\tau_{ei}(\epsilon)\rangle \\
C_0&=&\frac{2\surd{2}}{3\pi^2}
\end{eqnarray}
A generic form for $\tau_{ei}(\epsilon)$ can be obtained from the Rutherford formula
for the scattering of an electron by a heavy ion of charge $\bar{Z}$. This leads to the
Landau-Spitzer form if written in terms of a Coulomb Logarithm $C_{lg}$ (see
Sec 44, of ~\cite{LLvol10}).
Here we have restored the constants $m_e$ and $e$ for clarity.
\begin{eqnarray}
\label{tau_e.eqn}
\tau_{ei}(\epsilon)&=&\frac{m_e^{1/2}(2\epsilon)^{3/2}}{4\pi Z e^4\bar{n}C_{lg}}\\
                   &=& C_{ei}\epsilon^{3/2}
\end{eqnarray}
A more sophisticated calculation of the scattering, inclusive of the ion-distribution
by including a structure factor, pseudopotentials or a T-matrix usually amounts to an
improved form for $C_{lg}$. Our main purpose here is to provide a tractable form
for $\tau_{ei}(\epsilon)$ to evaluate the kinetic coefficients analytically, and the 
analysis remains valid as long as any improved from for $C_{lg}$ does not introduce any
additional dependence on $\epsilon$. Then, for $T>0$, the $df/d\epsilon$ can be reduced
by a partial integration to give:
\begin{eqnarray}
\label{kincoefs.eqn}
\mathcal{L}^{(0)}&=&C_1\int d\epsilon f(\epsilon)3\epsilon^2, \; C_1=C_0C_{ei} \\
\mathcal{L}^{(1)}&=&C_1\int d\epsilon f(\epsilon)(4\epsilon^3-3\mu\epsilon^2)\\
\mathcal{L}^{(2)}&=&C_1\int d\epsilon f(\epsilon)(5\epsilon^4-8\mu\epsilon^3+3\mu^2\epsilon^2)
\end{eqnarray}
These results can be incorporated into the expression for the Lorentz number $L_N$.
\begin{eqnarray}
\label{L_N.eqn}
L_N&=&\frac{1}{T^2}\left[\frac{\mathcal{L}^{(2)}}{\mathcal{L}^{(0}}-
\left(\frac{\mathcal{L}^(1)}{\mathcal{L}^{(0)}}\right)^2\right]\\
   &=&\frac{5I_4(\eta)}{3I_2(\eta)}-\left(\frac{4I_3(\eta)}{3I_2(\eta)}\right)^2 \\
I_n(\eta)&=&\int_0^\infty dx\frac{x^n}{1+\exp(x-\eta)},\;\eta=\mu/T
\end{eqnarray}
Thus $L_N=L_N(\eta)$ is dependent only on $t=T/E_F$ since $\eta$, the reduced
chemical potential depends only on $t$. Furthermore, in a DFT-implementation,
$\mu$ is the {\it non-interacting} chemical potential $\mu^0$ of Kohn-Sham electrons.

The reduction of $\mathcal{L}^{\alpha}$ given above is not applicable in the
 limit $T=0$ when $df/d\epsilon$ reduces to a delta-function. Then the kinetic
coefficients containing any $(\epsilon-\mu)$ factors reduce to zero. Consequently
 a Sommerfeld expansion about $\epsilon=E_F$ is needed. In the limit $T\to 0$
it can be shown that:
\begin{equation}
\label{kappaT0.eqn}
\kappa =(1/3)C_e<V^2>\tau_{ie}, \;\;  t\sim 0
\end{equation}
Here $C_e$ is the electron specific heat per particle, while  $<V^2>$
is an electron mean square-velocity evaluated within the thermally smeared
scattering region $ E_F\pm T$ in $k$-space enclosing  the
Fermi energy. We take this to be
\begin{eqnarray}
<v^2>&=&\langle V^2\rangle>/\langle 1 \rangle>\\
    &=&6T\frac{I_{1/2}(\eta)}{I_{-1/2}(\eta)}
\end{eqnarray}
The electron specific heat is approximated from the temperature derivative of the
total internal energy $E_e=E_0+E_{xc}$ of the uniform electron
fluid (UEF) at the $r_s$ and $T$ corresponding to the $\bar{Z}$ of the material studied.
Thus, for $l$-Al at 2.35g/cm$^3$ at the melting point $\sim$ 933 K, 
$\bar{Z}=3$ and $r_s$ = 2.171, the case studied by Recoules et al~\cite{Recou05}. 
However, we study Al at 2.7 g/cm$^3$, $r_s$=2.07322.

The ideal UEF energy $E_0$ and $C^0_e$ are  easily calculated, while $E_{xc}$
 is available from several analytic
 models~\cite{PDWXC,PDWXC84}, as well as empirically from numerical 
simulations~\cite{Brown2013}. The simulation data have been  parametrized for the
free energy $F(r_s,t)$~\cite{KSDT14,Dornheim18}. The second $T$-derivative of the
 parametrized $F(r_s,t)$  is needed for $C_e$. Consequently, artifacts of the
 parametrization may affect the calculated $C_e$~\cite{Karasiev19}. 

However, Eq.~\ref{kappaT0.eqn} is applicable only essentially at $T=0$ and the
problems with the finite-$T$ XC-parametrization arise only for $t$ well beyond
 the regime of validity of Eq.~\ref{kappaT0.eqn}. The only result that we use
 from Eq.~\ref{kappaT0.eqn} is the limiting value of $L_N$ at $T=0$ that is used
 in the fit function that extends the domain of Eq.~\ref{L_N.eqn} to $T=0$ as well.

The approach used here can also be used to obtain the thermoelectric coefficient
as well. However, we will not present  calculations of these other
transport coefficients. Furthermore, the
electron XC-effects, embedding-energy effects etc.,
 neglected here would be treated in a separate study.

\section{Transport coefficients and electron-electron interactions}
The total Hamiltonian of a system of electrons and ions that we consider
can be written in a self-evident notation as:
\begin{equation}
H=\sum_s H_0^s + H_{ii}+H_{ei}+H_{ee}, 
\end{equation} 
The ideal terms $H_0^s, s=i,e$ contain the kinetic energy of non-interacting particles of
type $s$. In the type of systems that we consider in this study, the electric current
and the heat current are carried by the electrons, as the ions are treated mainly as 
heavy scattering centers that provide resistance to electron flow under the
applied gradients of temperature or electric potential. If we consider the calculation
of the electrical conductivity, this can be done via the Boltzmann equation, or via the
current-current correlation function of Kubo theory. 

While the scattering of electrons
from heavy, essentially  static ions is easily addressed by these theoretical methods, the
effect of scattering of electrons, and how they affect the electrically conductivity (and
other transport coefficients) are more complex. The collision frequencies
$\nu_{ei}$ and $\nu_{ee}$ are assumed separable and are usually evaluated
independently in such treatments. 
Systems where $\nu_{ee}$ are neglected are referred to as ``Lorentz plasmas''. 
Here we argue that DFT provides a means of side-stepping this problem by
mapping any election-ion  plasma to an equivalent Lorentz plasma. The two-body e-e
interaction, $H_{ee}$, is replaced by a one-body XC-correlation functional. 

Usually the electron distribution function $f(k)$ perturbed by the
electric field to $f(k)=f_0(k)+\delta f(k)$ is considered. Here $\delta f(k)$ is small
and linear in the applied (very weak) field. The effect of $H_{ee}$ enters
into transport coefficients via the modification of the screening function (e.g., from
the Lindhard function to RPA and beyond) contained in the scattering cross section,
 and by its effect on $\delta f(k)$. The heat current is additionally modified  by
 the effect of e-e interactions on the electron  specific heat. These quantities are
 evaluated to some order in perturbation theory by traditional treatments
 of distribution functions, quantum Green's functions, diagrammatic methods etc.,
 in dealing with $H_{ee}$. 

The difficulties and uncertainties inherent in this process may be understood by examining
the inclusion of e-e interactions in the dielectric function, or equivalently, in the
response function $\chi(k,\omega)$ of the uniform electron fluid at finite-$T$, or even at
$T=0$. A treatment using the two-temperature Zubarev Green's functions has been given by
 the present author~\cite{Diel-CDW-76}. A major problem in these approaches is to obtain a 
conserving approximation, in the sense that the Ward identities, Gauge invariance etc.,
 should be obeyed by the approximation. Richardson and Ashcroft~\cite{RichAsh94} provided
 such a  finite-$T$ calculation to second order in the screened interaction. Applying the
 technique to the case of partially degenerate hydrogen plasma~\cite{CDW-Physica78} leads
 to  results which are extremely difficult to compute. It is difficult to ascertain if
 existing quantum-kinetic  results for e-e corrections to transport coefficients are
 conserving approximations.

The advent of density functional theory has provided an elegant solution to this
problem. DFT shows that the two-body $H_{ee}$ may be replaced by a one-body XC-functional
where the interacting electron gas is mapped to an equivalent non-interacting electron
system at the interacting-fluid density.
\begin{equation}
\label{DFT-H.eqn}
H=(H_0^i + V^i_{xc}([\rho]))+ (H_0^e+V^e_{xc}([n])+ H_{ei} 
\end{equation}
Here the two-body ion-ion interaction has been replaced an ion-XC potential~\cite{DWP82}.
It is this $V^i_{xc}([\rho])$ that enables us to use one-ion DFT, viz., the NPA, instead
 of the $N$-ion DFT used in QMD. The two-body e-e interaction is replaced by an
 electron-XC potential $V^e_{xc}([n])$ for which many finite-$T$ parametrizations are
 available. In Eq.~\ref{DFT-H.eqn} the
original Hamiltonian is reduced to that of a Lorentz plasma with no two-body
e-e scattering. Thus, if transport coefficients are calculated using DFT-generated
 cross sections, distribution functions etc.,
then no $\nu_{ee}$ contributions need to be included, although such
corrections may be needed in non-DFT theories of Spitzer and H\"{ar}rm~\cite{SpHarm53},
Reinholtz et al~\cite{Reinholz2015}  and others.

In our NPA calculations, the scattering cross section is expressed either in
terms of  a screened pseudopotential $U_{ei}(k)/\varepsilon(k)$, or via a T-matrix.
The pseudopotential is $u_{ei}(k)=\Delta n_f(k)/\chi(k)$, where $\chi(k)$ is the
electron response function. It is for an ion with an effective charge $\bar{Z}=n_f$
and a rigid core of bound electrons with $n_b$ electrons. The nuclear
charge of the ion $Z_n=n_b+n_f$. The electron XC functional enters into the determination
of $n_f(r)$ and hence into all the distribution functions. 

Similarly, the T-matrix provides a scattering cross section which involves
 the phase shifts that result from interactions with the nucleus as well
as all the electrons, bound and free, via the electron XC potential as
well as the Coulomb interactions. The collision 
frequency $\nu_{ei}$ calculated via either the NPA $U_{ei}(k)$, or
via the NPA generated T-matrix is for Kohn-Sham electrons constituting
a Lorentz plasma which {\it already incorporates the e-e collisions} in a
non-factorizable way, and to all orders in the e-e interaction.

\section{The electrical conductivities from the T-matrix and from $U_{ei}(k)$}
\begin{figure*}[t]
\includegraphics[width=1.8\columnwidth]{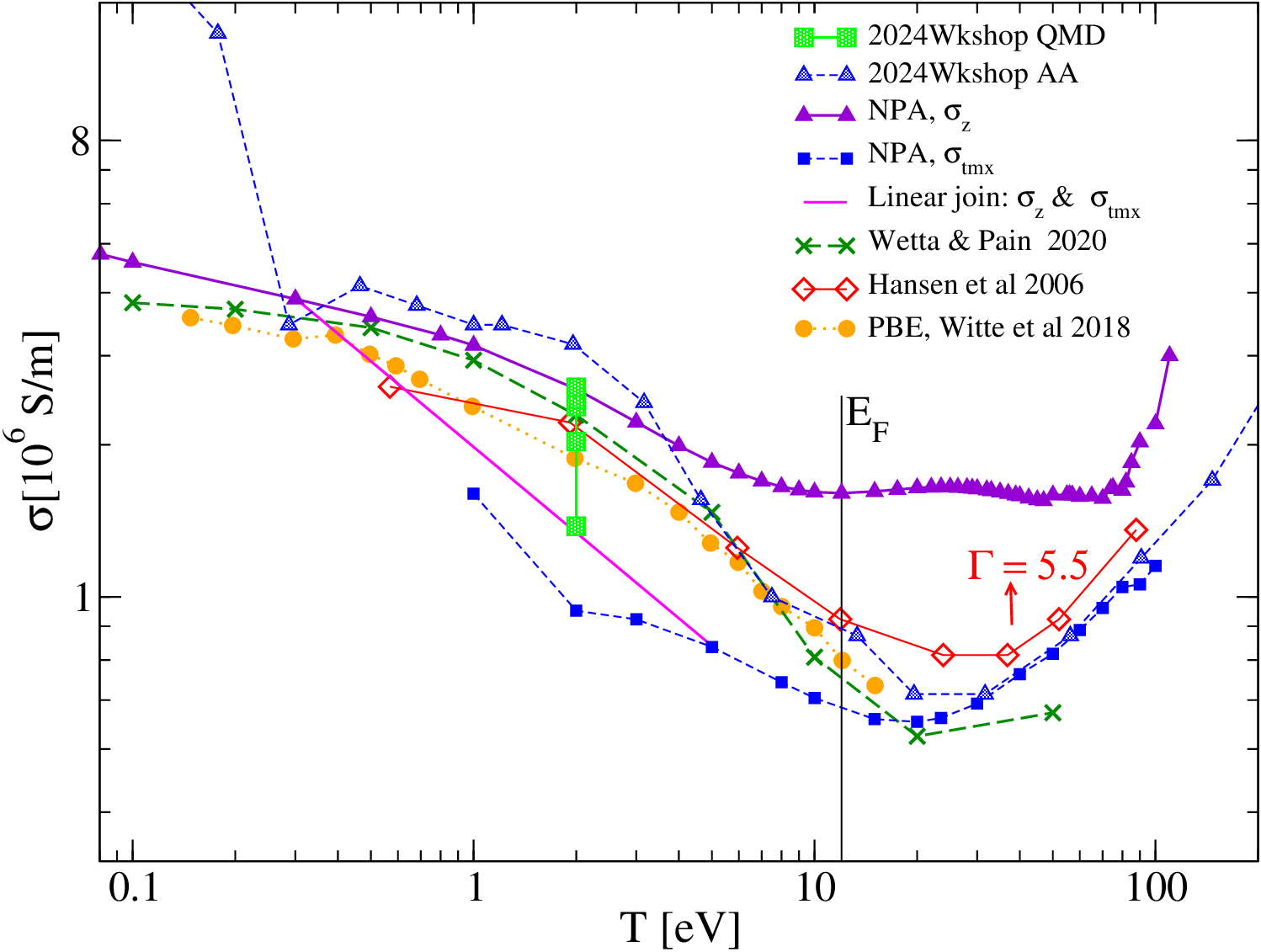}
\caption{\label{Alsig.fig}(Color online) The isochoric conductivities
$\sigma_z,\sigma_{tmx}$  calculated via the Ziman formula with NPA inputs, for
 $l$-aluminum, are compared with representative AA, QMD and other results. Results
given in Ref.~\cite{Stanek24} are labeled ``2024 Wkshop''.The five QMD
results at 2 eV use several different XC-functionals. Witte et al,
 2018~\cite{WittePOP18} refers to QMD calculations using the $T=0$
 Perdew-Burke-Ernzerhof XC-functional. Wette \& Pain 2020~\cite{Pain20};
Hansen et al 2016~\cite{Hansen2006}.
The value of $E_F$ displayed
corresponds to $\bar{Z}\sim 3$, but this increases as $T$ increases, as seen in
Table~\ref{Al-tszk.tab}.
}
\end{figure*}

In Fig.~\ref{Alsig.fig} we have displayed a number of calculations of the 
isochoric conductivity of $l$-aluminum at 2.7 g/cm$^3$.

As first noted by Perrot and the present author in 1999~\cite{Thermophys99},
these results confirm that $\sigma_z$ from pseudopotential-based calculations,
and $\sigma_{tmx}$ from T-matrix based calculations, differ significantly.
 The differences appear when  Al$^{3+}$ begins to loose core electrons,
 with $\bar{Z}$ increasing beyond three.
 The Al$^{3+}$ ion has a robust filled core with the electronic
 configuration: $1s^22s^23p^6$. An electron
moving under an applied field will scatter from it elastically, with no interaction
with the core except for a form factor already included in the weak pseudopotential.
Interactions between continuum electrons and core electrons are possible but
these are not elastic collisions (where the energy change $\omega \ne 0$).
\begin{eqnarray}
U_{ei}(k)&=&\Delta n_f(k)/\chi(k)\\
\chi(k)&=&\chi^0(k)/\left[1+v_k(1-G_k)\chi^0(k)\right]
\end{eqnarray}
Here $v_k=4\pi/k^2$, and $\chi^0(k)$ is the Lindhard function for non-interacting
 electrons.  The local-field correction $G_k$
contains XC-corrections. The use of the pseudopotential corresponds to the
use of the Hamiltonian
\begin{equation}
\label{pseudoH.eqn}
H=H_0+\sum_{\vec{k},\vec{k}_1,\vec{k}_2} U_{ei}(k)A^{\dagger}_{\vec{k}_1}a^{\dagger}_
{(\vec{k}_2+\vec{k})}
a_{\vec{k}_2}A_{(\vec{k}_1+\vec{k})} + \mbox{other terms}
\end{equation}
Here $A^{\dagger}_{\vec{k}},A_{\vec{k}}$ are creation and annihilation operators
for ions, while $a^{\dagger}_{\vec{k}},a_{\vec{k}}$ are for electrons. The matrix
elements are written to indicate momentum conservation to be formally exact,
 but this is irrelevant for massive ions usually treated in the Lorentz
plasma model;  the momenta of 
ion-density fluctuations  
$\rho^{\dagger}_k=\sum_{\vec{k}_1}A^{\dagger}_{\vec{k}_1+\vec{k}}A_{\vec{k}_1}$ are not 
conserved unless ion dynamics is included.

If the ion core is robust, and if $U_{ei}(k)$ is weak, as is the case for Al$^{3+}$,
multiple scattering effects, strong-collisions etc., are negligible and the T-matrix
results should agree with those from the weak pseudopotential. Numerical limitations
in our codes prevent us from extending the T-matrix calculation of the conductivity to
low temperatures to verify this explicitly. In fact, our $\sigma_{tmx}$ for Al becomes
increasingly inaccurate for $T<3$ eV. In fig.~\ref{Alsig.fig} we have joined the
$\sigma_z$ with $\sigma_{tmx}$ with a straight line to indicate the transition region
where the pseudopotential model begins to breakdown, while the T-matrix method
becomes appropriate. In this region core states acquire partial occupancies while
$\bar{Z}>3$ has a fractional value and an integer part.

The partial occupancies in the core provide a mechanism for some of the conduction
electrons to become ``hopping electrons''~\cite{hop1992}, and the value of the
 conductivity depends
on how these electrons are treated in the conductivity model. Partial occupancies
of core states make it possible for continuum electrons to interact with core
electrons while the overall energy is conserved, as in
elastic collisions, while the momentum need not be conserved as the ions are assumed
to be infinitely heavy. For instance, an electron in a $k,l$ state of energy $k^2/2$
may fall into a partially occupied $3p$ state while an electron in such a $3p$ may
be ejected to a $k,l'$ state of energy $k^2/2$. That is, the T-matrix approach includes
additional scattering channels that are not included in the ion with a rigid-core 
implied by the pseudopotential $U_{ei}(k)$.  

The phase shifts that are used to construct the T-matrix are such that:\\
 (i) they satisfy the finite-$T$ Friedel sum rule~\cite{DWP82} that sets
 the value of $\bar{Z}$ self-consistently
 with the ionization balance and thermodynamics; \\
(ii) they provide a consistent treatment of strong collisions that takes account
 of the partially ionized states of the core and any continuum resonances.  

Given that the numerical results of the pseudopotential model differ very significantly
from the strong-collisions model already at, say, $T=2$ eV, one would wonder why the
pseudopotential model is successful in accurately predicting the pair-distribution
functions of $l$-Al at 1 eV, or 2 eV, etc., in the sense that the $g(r), S(k)$ obtained
from the NPA pair-potentials agree very well from QMD calculations. The agreement
of NPA pair-distribution functions with those of QMD has been demonstrated in many
previous publications (e.g.,~\cite{HarbourDSF18, DW-yuk22}).
 The reason for this is that the ion-ion pair potential involves a strong repulsive
 term $\bar{Z}^2V_k$, which is not there in the electron-ion
interaction. The latter is essentially an attractive interaction that encourages
close collisions; furthermore, any Pauli blocking that
exists in fully occupied core states is removed for partial occupancies.

These same issues affect the accuracy of the Kubo-Greenwood approach in calculating
a $\sigma(\omega)$ and extrapolating to $\omega\to 0$ via some rigid-core model, e.g.,
the Drude model. The sensitivity of $\sigma_{K-G}$ obtained from QMD to the
 XC-functionals emphasizes this difficulty. In Fig.~1, the QMD K-G $\sigma$ for Al
at 2.7g/cm$^3$ takes the highest value of 2.6$\times 10^6$ S/m for a calculation
using the PBE functional, while the lowest value is nearly half, viz.,
 1.38$\times 10^6$ S/m is for the SCAN functional. It should be noted that as the number
$N$ of ions used in a QMD simulation increases, the complex character of possible 
ionic configurations of ``bonding schemes'' increases, and the corresponding electron
 distributions become very complex, demanding more and more complex XC-functionals. 
In contrast, in the NPA and in AA models there is only one ion and the corresponding electron
 density is a simple smooth density with the main rapid changes and discontinuity
 being at the nucleus. Consequently, NPA calculations are  insensitive to the
 XC-functional used.

Furthermore, since QMD implementations do not usually
incorporate finite-$T$ XC-functionals, the corrections to the specific heat from XC-effects
are not included in such calculations, thus affecting the calculation of $\kappa$.

\begin{figure}[t]
\includegraphics[width=0.96\columnwidth]{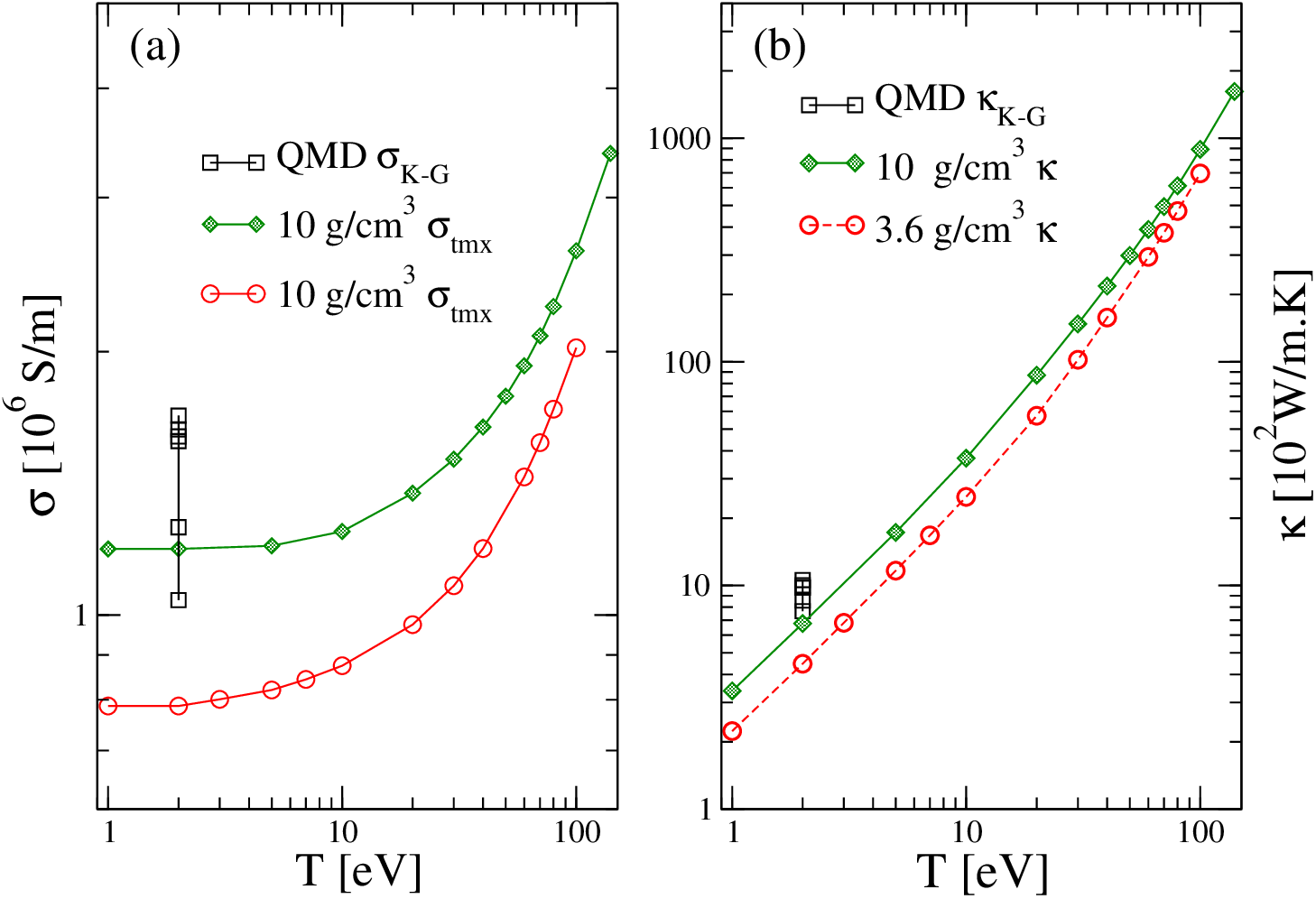}
\caption{\label{sig10-3p6Carb.fig}(Color online) (a) The isochoric conductivities 
$\sigma_{tmx}$ for $l$-carbon at 10 g/cm$^3$ and 3.6g/cm$^3$ from NPA calculations.
(b) The corresponding NPA-$\kappa$ obtained via the Lorentz number $L_n(t)$. 
The QMD results for $\sigma$ and $\kappa$ for $l$-C at 10 g/cm$^3$, $T$=2 eV 
reported in Ref.~\cite{Stanek24} are also displayed. 
}
\end{figure}

\section{Tabulated data for Aluminum and Carbon}
In this section we provide some representative results for isochoric 
$\sigma$ and $\kappa$ for $l$-carbon, and $l$-aluminum at 2.70 g/cm$^3$.
The $l$-carbon data are at the
density 10.0 g/cm$^3$ and at the ``diamond-like'' density of 3.6 g/cm$^3$.
 The $\sigma_{tmx}$ is calculated using the NPA, and $\kappa$ is obtained
 from the numerical fit to the Lorentz number $L_N(t)$.

A sample of tabulated results for $l$-carbon at 10 g/cm$^3$ is give in
table~\ref{C10-tszk.tab}

\begin{table}[h]
\caption{ \label{C10-tszk.tab} The  isobaric conductivity
$\sigma_{tmx}$,
 and $\kappa$ for $l$-C at 10 g/cm$^3$. The
thermal conductivity is calculated from $\sigma$ using the
finite-$T$ Lorentz number $L_N(t)$ defined in the main text.
More digits than warranted by physical accuracy are shown for
technical reasons (e.g., useful in identifying the version of
a code used to generate the results).
}
\begin{ruledtabular}
\begin{tabular}{lccc}
Tev &$\sigma_{tmx}$ & $\bar{Z}$  & $\kappa/10^2$ \\
\hline\\
1    &  1.190     &  4.000     &  3.373 \\
2    &  1.190     &  4.000     &  6.743 \\
5    &  1.191     &  4.000     &  16.89 \\
10   &  1.246     &  4.000     &  35.34 \\
20   &  1.378     &  4.000     &  78.40 \\
40   &  1.640     &  4.003     &  193.4 \\
60   &  1.927     &  4.042     &  365.2 \\
80   &  2.252     &  4.158     &  595.4 \\
100  &  2.608     &  4.337     &  879.6 \\
140  &  3.369     &  4.733     &  1613 \\
\end{tabular}
\end{ruledtabular}
\end{table}

At low temperatures, the resistivity ``saturates'' as the $l$-carbon
structure factor adjusts to have a maximum at 2$k_F$ as
$T/E_F\to 0$, when electron-ion scattering is maximized
as in Friedel-controlled fluids~\cite{DW-yuk22}.
We see essentially the same behaviour in 
 carbon at the ``diamond-like'' density of 3.6 g/cm$^3$,
(see Table~\ref{C3p6g-tszk.tab}). The data for $l$-carbon
at these two densities are displayed in Fig.~\ref{sig10-3p6Carb.fig}.

\begin{table}[h]
\caption{ \label{C3p6g-tszk.tab} The  isobaric conductivity
$\sigma_{tmx}$, and $\kappa$ for $l$-C at 3.6 g/cm$^3$. The
thermal conductivity is calculated from $\sigma$ using the
finite-$T$ Lorentz number $L_N(t)$ defined in the main text.}
\begin{ruledtabular}
\begin{tabular}{lccc}
Tev &$\sigma_{tmx}$ & $\bar{Z}$  & $\kappa/10^2$ \\
\hline\\
    1.0 &  0.787 &  4.000  &   2.233 \\
    2.0 &  0.787 &  4.000  &   4.464 \\
    3.0 &  0.801 &  4.000  &   6.813 \\
    5.0 &  0.821 &  4.000  &   11.64 \\
   10.0 &  0.875 &  4.000  &   24.89 \\
   20.0 &  0.975 &  4.000  &   57.41 \\
   40.0 &  1.191 &  4.005  &   157.6 \\
   60.0 &  1.438 &  4.073  &   294.5 \\
   80.0 &  1.719 &  4.267  &   472.4 \\
  100.0 &  2.021 &  4.530  &   695.5 \\
\end{tabular}
\end{ruledtabular}
\end{table}

\begin{table}[h]
\caption{ \label{Al-tszk.tab} The  isobaric conductivity
$\sigma_{tmx}$ ($\sigma_z$ if in parenthesis),
 and $\kappa$ for $l$-Al at 2.7 g/cm$^3$. The
thermal conductivity is calculated from $\sigma$ using the
finite-$T$ Lorentz number $L_N(t)$ defined in the main text.
More digits than warranted by physical accuracy are shown for
technical reasons (e.g., useful in identifying the version of
a code used to generate the results).
}
\begin{ruledtabular}
\begin{tabular}{lccc}
Tev &$\sigma_{tmx}$ & $\bar{Z}$  & $\kappa/10^2$ \\ 
\hline\\
0.1 &      (4.60)   &   3.000    & 1.303 \\ 
0.3 &      (3.89)   &   3.000    & 3.309 \\
5   &      0.796    &   3.000    & 12.83 \\
8   &      0.678    &   3.003   & 15.98 \\
10  &      0.631    &   3.016   & 21.19 \\
20  &      0.566    &   3.495   & 38.62 \\
30  &      0.615    &   4.299   & 63.22 \\
40  &      0.703    &   5.085   & 96.55 \\
60  &      0.860    &   6.233   & 177.5 \\
80  &      1.046    &   7.296   & 288.2 \\
100 &      1.152    &   7.720   & 396.9 \\
\end{tabular}
\end{ruledtabular}
\end{table}

%